\documentclass[a4paper,fleqn,usenatbib]{mnras}

\newcommand{\E}{}
\newcommand{\mkm}{\mu m}

\usepackage{graphicx}	
\usepackage{amsmath}	
\usepackage{amssymb}	
\usepackage{multicol}
\usepackage{threeparttable}

\title[A0620-00 photometry in passive and active stages]{Optical and $J$, $K$-photometry of black hole X-ray nova A0620-00 in passive and active stages of quiescence}
\author[A.M. Cherepashchuk et al.]{A.M.~Cherepashchuk$^{1}$, N.A.~Katysheva$^{1}$, T.S.~Khruzina$^{1}$, S.Yu.~Shugarov$^{1,2}$,\newauthor A.M.~Tatarnikov$^{1}$, M.A.~Burlak$^{1}$,
N.I.~Shatsky$^{1}$\thanks{Contact e-mail: kolja@sai.msu.ru} \\ $^{1}$Lomonosov Moscow State University, Sternberg Astronomical Institute, Universitetsky prosp. 13, 119992, Moscow, Russia\\
$^{2}$Astronomical Institute, Slovak academy of sciences, 05960, Tatransk\'{a} Lomnica, Slovakia}

\begin{document}
\date{Submitted 2017 November 15}

\pagerange{\pageref{firstpage}--\pageref{lastpage}}
\pubyear{2017}
\maketitle
\label{firstpage}

\begin{abstract}
\E Photometric observations of the low-mass X-ray binary system A0620-00=V616~Mon are performed in the optical (unfiltered light, $\lambda_{\textrm{eff}}\approx6400$~\AA) and the near-infrared $J$ and $K$-bands. The mean system flux, the orbital light curve shape and the flickering amplitude dependences on wavelength are examined in the $\lambda\lambda 6400-22000$~\AA\ range for two activity stages of the system which remains in quiescence.

\E In 2015--16 the A0620-00 system was observed to be in {\em passive} stage (according to the terminology of Cantrell et al., 2008) exhibiting the regular orbital light curve with a relatively low flickering level.

\E In less than 230 days in 2016--17 the system switched into {\em active} stage: its mean brightness increased by $\sim 0.2-0.3$~mag in all the studied bands, the orbital light curve experienced drastic changes while the flickering amplitude increased more than twice both in the optical and in the infrared.

\E The regular orbital light curves of A0620-00 were modelled within two frameworks, namely assuming existence of ``cold'' spots on the optical companion surface and without those. These models allow for a good fit of observed orbital light curves both in passive and in active stages. A close correlation between the luminosity of accretion disc, ``hotline'' and the flickering amplitude is established.

\E The absolute fluxes having been calibrated and corrected for the interstellar extinction, the dependence of the mean square flickering amplitude (in fluxes) is computed as a function of wavelength in the $\lambda\lambda6400-22000$~\AA\ range.

\E In active stage, the observed flickering amplitude decreases as a function of wavelength over the whole studied range and may be adequately represented by a single dependence $\Delta F_{\textrm{fl}}\sim\lambda^{-2}$ which corresponds to the thermal free-free emission of optically thin high-temperature plasma.

\E In passive stage, the $\Delta F_{\textrm{fl}}$  value behaves like $\Delta F_{\textrm{fl}}\sim\lambda^{-4}$ in the short wavelength range of $\lambda\lambda6400 - 12500$~\AA\ that corresponds to the thermal radiation of optically thick high-temperature plasma. In the long-wave domain $\lambda\lambda12500 - 22000$~\AA\ the flickering amplitude dependence is flat: $\Delta F_{\textrm{fl}}\sim const$ which may imply existence of a synchrotron component of the relativistic jets emission.

\E The $\Delta F_{\textrm{fl}}$ dependence on the mean system flux in $K$-filter is shallower than corresponding relations for the filter $J$ and for the optical range.

\E The above mentioned flickering features let us propose that the mechanism responsible for flickering includes at least two components: thermal and, apparently, synchrotron, that agrees with the recent discovery of the variable linear polarization of the infrared system emission (Russell et al., 2016).
\end{abstract}


\section{Introduction}\label{sec:intro}
\E X-ray novae with black holes (BH) provide us the majority of information on the masses of stellar-mass black holes (see, e.g. \citealt{casares_jonker2014}). These are soft X-ray transients that experience powerful outbursts of X-ray and optical radiation which last for several months. In intervals between outbursts, a {\em quiescence} state is observed in the binary system which consists of a low-mass optical companion star and a relativistic object surrounded by an accretion disc. In this state, characterized by the X-ray luminosity $L_X < 10^{33}$~erg~s$^{-1}$, the principal mechanism  of the optical and infrared (IR) flux variability is the ellipticity of the optical star ({\em ellipticity effect}, \citealt{lut_sun_cher1973,lut_sun_cher1975}). As the luminosity of optical star in low-mass X-ray binaries is relatively low, the accretion disc and the region of its interaction with the infalling gas stream provide a significant part of emission, and so do the relativistic jets, whose synchrotron radiation causes the observed IR light polarization \citep{russell2006,russell2016}. That is why the regular orbital flux variations due to the ellipticity effect are modulated by different kinds of additional light variability induced by the gas structure changes. This pattern may be further complicated by effects of the optical star physical variability and modulated by spot structures on its surface (see e.g. \citealt{Khruzina_cher1995, zurita2016}).

\E The decades-long experience of the observational studies of X-ray binaries reveals the following types of additional flux variations which superpose over the orbital optical and IR light curves \citep{cantrell2008, casares_jonker2014}:

\begin{enumerate}
\E \item The so called {\em superhump} variability which affects the orbital light curve via the precession of the accretion disc semi-major axis happening during an outburst. This phenomenon may also be observed in quiescence if the contribution of the accretion disc and gas stream to the system total flux is relatively high in this state. As \citet{zurita2016} show, the accretion disc ellipticity may persist for as long as ten years after the outburst which witnesses a relatively low viscosity of the disc material in quiescence.
\E \item Fast aperiodic and quasi-periodic variability (so called {\em flickering}) caused by magneto-hydrodynamic processes going on in disc, in stream-disc interaction region and in jets.  For example, the photometric observations of the A0620-00=V616~Mon system in the period 2003--2016 by \citet{shugarov2015,shugarov2016} allow to trace the optical light curve evolution with time which is a combination of periods of relatively regular flux variations ({\em passive stage}, see below) and the ones when the ellipsoidal orbital light curve is overlaid by irregular flares (flickering), corresponding to the {\em active stage}.
\end{enumerate}

\E It is generally adopted to use the term {\em flickering} in a close binary system for fast aperiodic and quasi-periodic light oscillations with an amplitude from percents to tens of percents at the time-scales from seconds to tens of minutes. Flickering is commonly thought to be related to magneto-hydrodynamic processes in the accretion disc, in a disc--stream interaction region and perhaps some non-stationary processes in the relativistic jets. Meanwhile, the nature of this phenomenon is not yet well-established (see, e.g. \citealt{bruch2000, bruch2015, baptista2015}). Flickering is observed in dwarf novae, polars, symbiotic stars and X-ray binaries.

\E In recent years the flickering properties of accreting objects have been studied intensively. For example, \citet{zamanov2015} have traced flickering in the recurrent nova RS~Oph in quiescence in five photometric bands ($UBVRI$). Authors point out that the flickering amplitude increases when the mean flux of the system rises.

\E According to \citet{cantrell2008}, there are three stages of activity which may be distinguished during the quiescent state of a low-mass X-ray binary: passive, intermediate and active. As the object migrates from passive to active stage, the mean system flux goes up and the irregular variability amplitude increases. In  passive stage, the optical colour becomes, as a rule, redder and the flickering intensity is the lowest. This is why the ellipticity effect is thought to dominate during passive stage \citep{cantrell2008,cantrell2010}. Orbital inclination $i$ derived from the analysis of light curves in quiescence is considered the most reliable.

\E The A0620-00 system is observed since 1975 (more than 40 years) and by now is studied thoroughly. Nevertheless, in recent years principally new results on the fast secular orbit period evolution have been obtained. Moreover, the variable linear polarization in the infrared was discovered in the quiescent state. It leads to the necessity to continue intensive wide band photometric and spectral observations of A0620-00.

\E In this paper we examine the dependence of the mean system flux, orbital light curve shape and the flickering amplitude on wavelength in the range from 6400~\AA\ to $2.2~\mkm$ for two stages of activity in this system which up to now remains in the quiescent state.

\section{System description}\label{sec:sys_description}


\E The X-ray nova A0620-00 was discovered during an X-ray outburst  in 1975. The observed X-ray maximum flux was 50 times larger than the Crab nebula flux which made A0620-00 the brightest X-ray source in the sky at that moment \citep{elvis1975}. Examination of archive photographic data allowed to establish that related optical variable V616~Monocerotis underwent a nova outburst also in 1917 (see, e.g. \citealt{cher2000} and references therein).


\E After V616~Mon returned into quiescent state its X-ray luminosity fell down to $3\times10^{30}~\textrm{erg~s}^{-1}$  \citep{garcia2001} which constituted a small part of the optical component bolometric flux. The red dwarf of the spectral type K3--7V (according to different determinations, \citealt{Froning2007}) fills its Roche lobe. A0620-00 is the first low-mass X-ray binary system in which the relativistic component was identified with a black hole \citep{McClintock_Remillard}, because the optical component mass function was estimated as $f_\nu(M)=3.18 \pm 0.16~\textrm{M}_{\sun}$, well above the neutron star mass limit predicted by Einstein's General Relativity. This conclusion was later confirmed by a number of researchers \citep{Marsh1994,Orosz1994,Neilsen2008}. The Doppler tomography method revealed an accretion disc asymmetry and existence of a bright hotspot in the region of the gas stream interaction with the disc material. The disc luminosity is variable in quiescence and well noticeable both in optical and in the infrared. As noticed above, depending on the current disc state, the V616~Mon light curves exhibit three distinct stages which differ in brightness, colour and character of irregular variability \citep{cantrell2008}.


\E The system parameters were derived by many researchers (see, e.g. \citealt{casares_jonker2014} for a review of this problem), since even if the optical star mass function is given precise relativistic object mass determination depends on orbital inclination $i$ and the components mass ratio. In the absence of eclipses, different methods involved to constrain $i$ lead to results poorly corresponding to each other. The obtained values range from  $i=36\fdg6$ \citep{Shahbaz} to $75\degr$ \citep{Fron_Rob} depending on the observations epoch, the instant light curve shape and interpretation approaches. Resulting relativistic object mass varies between $3.4~\textrm{M}_{\sun}$ for $i=75\degr$ and $13.2~\textrm{M}_{\sun}$ for $i=37\degr$ \citep{Marsh1994}. A seemingly most reliable mass estimate was derived in the work of \cite{cantrell2010} who, involving numerous optical and infrared light curves of A0620-00,  extracted those curve patterns which correspond to passive system stage. So, by spectrophotometric methods they estimated the accretion disc contribution which turned out to be substantial even in the infrared.  Their result was the inclination $i=51\fdg0 \pm 0\fdg9$ and the BH mass $M_\textrm{X} = 6.6 \pm 0.25~\textrm{M}_{\sun}$, which may be accounted the most reliable to date. In the same paper the reference distance estimate for A0620-00 is given: $d=1.06 \pm 0.12$~kpc.


\E The discovery of anomalously fast orbital period decay for three X-ray novae with BHs А0620-00, XTE~J1118+480 = KV~UMa, Nova~Mus~1991 = GU~Mus \citep{GonHer2014, GonHer2017} as well as for the low-mass X-ray binary (LMXRB) with the neutron star AX~J1745.6-2901 \citep{Ponti} sets the task to clarify the optical donor stars nature for these systems. The period shortening rates are $\textrm{d}P/\textrm{d}t = -0.60\pm0.08$, $-1.90\pm0.57$ and $-20.7\pm12.7~\textrm{ms~yr}^{-1}$ for the systems with black holes А0620-00, XTE~J1118+480, Nova~Mus~1991, respectively. These rates are so huge that, given they are constant in time, the formal life time for these systems (a time span for optical stars to reach the BH event horizon) appears to be $70\times10^6$~yr for А0620-00, $12\times10^6$~yr for XTE~J1118+480 and $2.7\times10^6$~yr for Nova~Mus~1991 \citep{GonHer2017}. At the same time, according to the standard evolutionary scenario for LMXRBs in which the binary loses the orbital momentum via magnetic stellar wind from the optical star and gravitational waves radiation, the characteristic  orbital evolution time has to be quite considerable, around $5\times10^9$~yr \citep{Iben}.


\E Evolutionary scenarios for X-ray novae with BHs, both standard and non-standard, are discussed in papers of \cite{GonHer2014, GonHer2017}. Nevertheless, if one adheres the most natural standard scenario, it requires to propose huge magnetic field strengths up to tens of kilogauss \citep{GonHer2014}. It is natural to foresee a noticeable chromospheric activity on the optical companions of X-ray novae in these conditions, as well as intensified formation of spots in the photospheres of these stars.  Traces of chromospheric activity of optical companions have been recently discovered spectroscopically in А0620-00 and XTE~J1118+480 \citep{GonHer2010, zurita2016}. In works of \citep{Katsova, Bildsten} it is shown that coronal X-ray radiation of optical stars in LMXRBs of late spectral types may contribute considerably in the total X-ray flux from novae in quiescence, which gives reason to consider these companions as chromospherically active stars.


\E These circumstances force the necessity to take into account possible existence of spots on the surfaces of tidally deformed optical stars (as well as existence of an accretion disc with its interaction region with the gas stream) while interpreting the optical and IR light curves of LMXRBs. This was firstly performed in the work of \citet{Khruzina_cher1995} for interpretation of long-term orbital curves variability of the X-ray nova A0620-00 where it was shown that such a curves variation may be reconstructed by appearance, motion and dissolving of ``dark'' spots on the stellar surface. One of the goals of the current paper is indeed the account for spot structures of A0620-00 on the base of an advanced LMXRB system model (see, e.g. \citealt{Khruzina2005}) and new observations.


\E Another new important achievement of last years was the discovery of a linear polarization of the infrared radiation from LMXRBs with BHs Swift~J1337.2-0933 and А0620-00 in quiescence \citep{russell2016}. This was related to synchrotron radiation from relativistic jets which exist not only during X-ray outbursts but in quiescence as well. Polarization data evidence the degree of magnetic field perplexity to be nearly the same during an outburst and in a quiet state.


\E Given this recent discovery, the optical and IR observations of A0620-00 present a considerable interest for detailed research of flickering phenomenon and its dependence on wavelength which may shed light on physical mechanisms of flickering.


\E In the coming Section~\ref{sec:observations} we present the A0620-00 photometry data in the optical white light and in the $J$- and $K$-bands of the near-infrared. The following Section~\ref{sec:modelling} describes the system models and interpretation methods in application to its light curves. In Section~\ref{sec:mod_results} the results of interpretation are outlined for the optical star models both with spots and without those. The residuals analysis for observed light curves and flickering characterization are performed in Section~\ref{sec:6}. Results discussion and conclusions are given in Section~\ref{sec:conclusion}.

\section{Observations}\label{sec:observations}

\E Observations of А0620-00 = V616~Mon were performed during two winter seasons of 2015--16 and 2016--17 in optical and near-infrared ranges. Each night allowed to observe the system during approximately half of the orbital period. The observation log is given in Table~\ref{tab1}.

\begin{table*}
\caption{Log of optical and IR-observations for A0620-00.}
  \label{tab1}
\begin{threeparttable}
  \begin{tabular}{ccccccccc}
  \hline
  Date & JD  & \multicolumn{2}{c}{Filter and designation} & $N$ & $\varphi_{\textrm{i}}-\varphi_{\textrm{f}}$ & $m_{\textrm{min}}$ & $m_{\textrm{max}}$ & Telescope\\
 dd.mm.yyyy      & 2450000+ & \multicolumn{2}{c}{of observational epoch} & & & & & \\
       \hline
04.11.2015 & 7331.435--.653	& \multicolumn{2}{c}{} & 399 & 0.389--1.063 & 3.010 & 2.535 & 1.25~m\\
05.11.2015 & 7332.449--.653	& \multicolumn{2}{c}{} &	376	& 0.529--0.158 &	2.938 &	2.621 & 1.25~m\\
16.12.2015 & 7373.447--.528 & \multicolumn{2}{c}{} & 7 & 0.451--0.701 & 2.902 & 2.680 & 1.25~m\\
22.12.2015 & 7378.521--.544 & \multicolumn{2}{c}{$\Delta C15-16$} & 3 & 0.158--0.230 & 2.826 & 2.785 & 1.25~m\\
22.12.2015 & 7379.295--.531 & \multicolumn{2}{c}{} & 19 & 0.557--1.287 & 2.978 & 2.664 & 1.25~m\\
03.01.2016 & 7391.306--.591	& \multicolumn{2}{c}{} & 139 & 0.739--1.621 & 2.024 & 2.662 & 0.6~m\\
29.03.2016 & 7477.258--.366 & \multicolumn{2}{c}{} & 55 & 0.835--1.168 & 2.909 & 2.557 &	0.6~m\\
31.03.2016 & 7479.267--.334 & \multicolumn{2}{c}{} & 39 & 0.054--0.261 & 2.974 &	2.705 & 0.6~m\\
31.01.2016 & 7419.182--.346 & & $J419$ & 161 & 0.039--0.547 & 15.545 & 15.408 & 2.5~m\\
02.02.2016 & 7421.216--.408	& $J16$ & $J421$ & 182 & 0.335--0.932 & 15.527 & 15.366 & 2.5~m\\
03.02.2016 & 7422.233--.448 & & $J422$ & 201 & 0.486--1.151 & 15.537 & 15.334 & 2.5~m\\
31.01.2016 & 7419.188--.344 & & $K419$ & 154	& 0.058--0.540 & 14.608 & 14.440 & 2.5~m\\
02.02.2016 & 7421.221--.397 & $K16$ & $K421$ & 170 & 0.351--0.897 & 14.484 & 14.364 & 2.5~m\\
03.02.2016 & 7422.238--.453 & & $K422$ & 183 & 0.501--1.164 & 14.529 & 14.331 & 2.5~m\\
19.11.2016 & 7711.503--.662	& \multicolumn{2}{c}{}& 236 & 0.020--0.509 & 2.717 &	2.204 & 1.25~m\\
19.11.2016 & 7712.475--.659 & \multicolumn{2}{c}{} & 342 & 0.028--0.510 & 2.770 & 2.229 & 1.25~m\\
20.11.2016 & 7713.400--.661 & \multicolumn{2}{c}{} & 455 & 0.892--1.700 & 2.930 & 2.273 & 1.25~m\\
21.11.2016 & 7714.430--.514 & \multicolumn{2}{c}{} & 141 & 0.080--0.341 & 2.753	& 2.237 &	1.25~m\\
22.11.2016 & 7715.436--.658 & \multicolumn{2}{c}{} & 388 & 0.194--0.883 & 2.834 & 2.347 & 1.25~m\\
30.11.2016 & 7723.441--.546 & \multicolumn{2}{c}{$\Delta C16-17$} & 284 & 0.976--1.301 & 2.925 & 2.221 & 1.25~m\\
05.12.2016 & 7728.455--.549 & & & 136 & 0.500--0.791 & 2.815 & 2.266 & 1.25~m\\
23.01.2017 & 7777.264--.493 & & & 159 & 0.603--1.313 & 2.856 & 2.402 & 0.6~m\\
26.01.2017 & 7780.344--.527	& & & 127 & 0.139--0.704 & 2.807 & 2.379 & 0.6~m\\
27.01.2017 & 7781.294--.421	& & & 70 & 0.079--0.475 & 2.870 & 2.502 & 0.6~m\\
28.01.2017 & 7782.290--.326	& & & 36 & 0.163--0.277 & 2.727 & 2.520 & 0.6~m\\
16.02.2017 & 7801.189--.414	& & $J801$ & 213 & 0.672--1.370 & 15.492 & 14.982 & 2.5~m\\
19.02.2017 & 7804.220--.408	& $J17$ & $J804$ & 190	& 0.056--0.636 & 15.351 & 15.599 & 2.5~m\\
06.03.2017 & 7819.163--.352 & & $J819$ & 203 & 0.317--0.901 & 15.388 & 16.061 & 2.5~m\\
16.02.2017 & 7801.194--.410 & & $K801$ & 202	& 0.688--1.355 & 14.240 & 14.007 & 2.5~m\\
19.02.2017 & 7804.225--.413	& $K17$ & $K804$ & 190 & 0.071--0.652 & 14.198 & 14.018 & 2.5~m\\
06.03.2017 & 7819.148--.337 & & $K819$ & 210 & 0.272--0.856 & 14.312 & 14.054 & 2.5~m\\
\hline
\end{tabular}


\E Notes: Date is UT date for the beginning of observations, JD denotes the observational interval during current night, $\Delta C15-16$ -- magnitudes difference of A0620-00 and the main comparison star (in white light) measured in 2015--16 (no interstellar absorption is taken into account), $J16$, $K16$ -- infrared $J$ and $K$ band target magnitudes in 2015--16 seasons. $\Delta C16-17$ and $J17, K17$ -- the same values for season 2016--17; $N$ -- number of obtained frames; $\varphi_{\textrm{i}},\varphi_{\textrm{f}}$ -- orbital phases of beginning and end of observations; $m_{\textrm{min}}, m_{\textrm{max}}$ -- minimal and maximal stellar magnitude during observations. Individual data points will be published later and are available upon request.

\end{threeparttable}
\end{table*}

\subsection{Near-IR observations}
\label{obs:ir}

\E Infrared observations in the bands $J$ ($\lambda_{\textrm{eff}}\sim1.25~\mkm$) and $K$ ($\lambda_{\textrm{eff}}\sim2.2~\mkm$) of the MKO photometric system were performed at the newly installed 2.5-m telescope of Caucasian highland observatory of SAI MSU, at elevation of 2112~m a.s.l. The site started operating in 2015 and was described in \citet{Sad_Cher} while its astroclimatic parameters were analysed in the paper of \citet{Kornilov}.



\E For IR observations the camera-spectrograph ASTRONIRCAM \citep{Nadjip} with the Hawaii-2RG detector of $2048\times2048$ pixels format with a set of wide-band filters was used. The photometric mode of operation was used where the central detector area of $1024\times1024$~pix is illuminated through $J,K$-filters. The image plate scale is $0.27~\textrm{arcsec~pix}^{-1}$, the field of view is a $4\farcm6$ side square. Monitoring was performed with filters interchange after each 9 or 10 frames of 29.17~s integration time during three nights in 2016 (Jan 31, and Feb 2, 3) and three nights in 2017 (Feb 16, 19, and Mar 3).

\begin{figure}
 \includegraphics[width=\columnwidth]{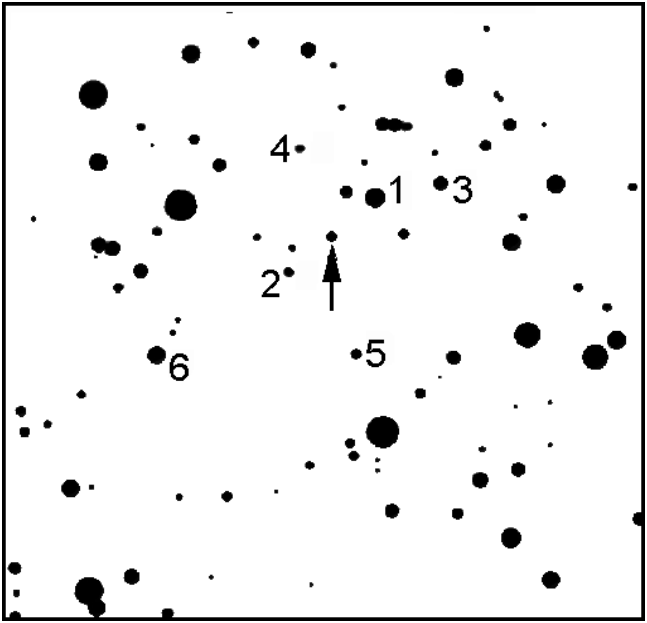}
 \E \caption{The V616~Mon field map. 1 -- primary comparison star (2MASS J06224334-0020262), 2--6 --- control stars.}
 \label{fig1}
\end{figure}

\E Resulting frame images were measured by the aperture photometry method with the primary comparison star 2MASS J06224334-0020262 which magnitudes were taken from the 2MASS catalogue and converted into the MKO system according to \citet{leggett} recipe ($J_{\textrm{MKO}}=13.247$~mag, $K_{\textrm{MKO}}=12.74$~mag). The photometry precision was controlled using other comparison (control) stars close to the object both in magnitude and position (marked with numbers 2--6 in Figure~\ref{fig1}). No brightness correlation between comparison stars and the target was observed; the root-mean-square (rms) error of the comparison star measurements varied between $0\fm015$ and $0\fm025$ depending on a particular date, so a value of $0\fm02$ may be taken as a mean rms error estimate for our individual $J,K$ band measurements.


\E The measured magnitudes were averaged within each filter series (by 3--9 points depending on observation conditions), thus yielding time sampling and photometric precision similar to those in the optical (see below). Such an averaging is believed to have no smoothing effect on measurements of flickering amplitudes since characteristic time of flickering for A0620-00 is of the order of 30--40~min \citep{shugarov2015,shugarov2016}. Thus we obtained 125 averaged infrared brightness estimates in 2016 and 434 estimates in 2017 which are used in the analysis presented in this paper.

\subsection{Optical observations}

\E Optical observations in white integral light (without filters, $\lambda_{\textrm{eff}}\sim6400$~\AA, the bandpass half-response wavelengths are $\lambda\lambda4300-8300$~\AA, see e.g. \citet{khruzina2015,armstrong2013} on the band discussion)  were taken at 1.25-m telescope of the Southern station of SAI MSU (CCD-camera VersArray-1300B, exposure times 45--60~s) and at the 0.6-m telescope of Slovak academy of sciences (Star\'a Lesn\'a, Slovakia; CCD-camera FLI ML 3041, exposure time 120~s). The object was observed on 2015 Nov 4, 5,  and Dec 16, 22; 2016 Jan 3, Mar 29, 31, Nov 19--21, 30, and Dec 5; 2017 Jan 23, 26--28.


\E Images taken in white light were processed with the same comparison and control stars which were used for IR observations (Section~\ref{obs:ir}). The primary comparison star magnitudes are $B=15\fm72$, $V=14\fm92$ and $R=14\fm2$ (see the catalogue of \citealt{cher1996}). The photometric intrinsic precision in white light (integral light, ``Clear filter = $C$'') is about $\sim0.01$~mag which was assessed using rms deviation of magnitude differences with the comparison stars, similarly to IR measurements. The number of independent magnitude estimates in white light was 1037 in the period from November 2015 to March 2016 and 2374 estimates for the time span from November 2016 to January 2017.

\subsection{Light curves}

\E The following ephemeris was employed for folding data into light curves:
\begin{equation}
\label{eq:ephem}
   T(\textrm{Min~I}) = \textrm{JD}~2457332.601507 + 0\fd32301407\times E
\end{equation}


\E The phase $\varphi=0.0$ for this ephemeris corresponds to the upper conjunction of the relativistic object (optical star is in front of the relativistic object). The initial phase in equation \ref{eq:ephem} is shifted by  $0.5P_{\textrm{orb}}$ with respect to ephemeris of \citet{shugarov2016} corresponding to the compact object being in front of the optical star.

\begin{figure}
 \includegraphics[width=\columnwidth]{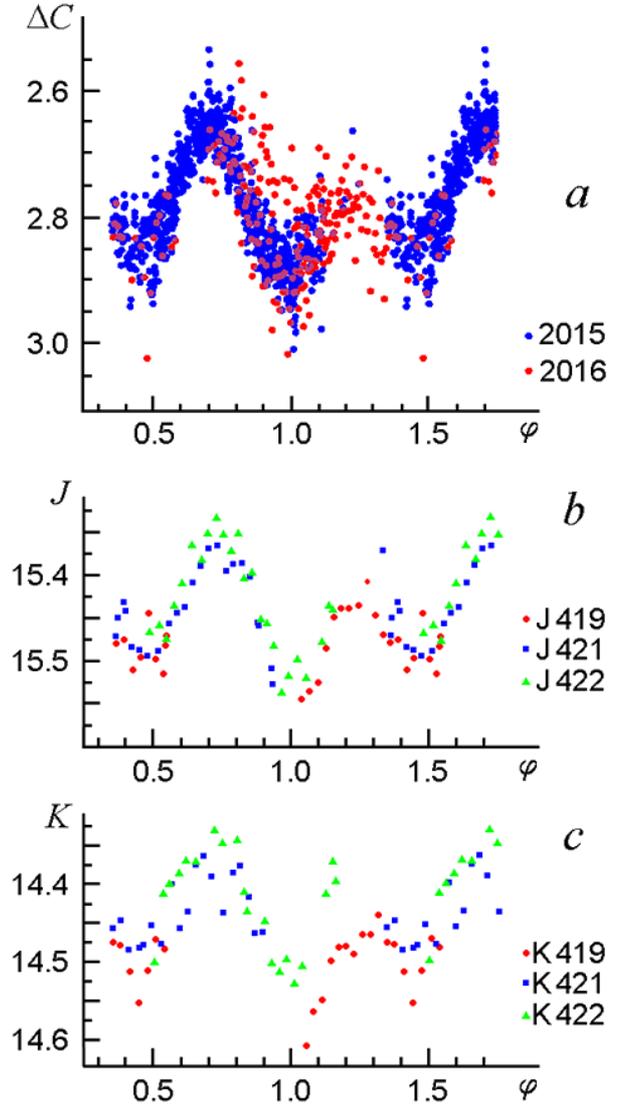}
 \E \caption{Photometric data on A0620-00 convolved with orbital period. From top to bottom: observation in ''Clear filter'' in 2015--2016, in $J$ and $K$ bands in 2016. The system was in passive stage.}
 \label{fig2}
\end{figure}

\E In Figure~\ref{fig2} the A0620-00 photometric measurements in white light ($\Delta C$ is the difference with the primary comparison star), as well as in bands $J$ and $K$, are given for the first season 2015--16 . The light curves for this season exhibit regular orbital variability with relatively small random deviations.

\E The light curve in white light for one orbital period represents a double wave  with unequal maxima, with a $\varphi=1$ minimum being deeper than that at $\varphi=0.5$. This property is evident in the IR bands $J$ and $K$ as well.

\begin{figure}
 \includegraphics[width=\columnwidth]{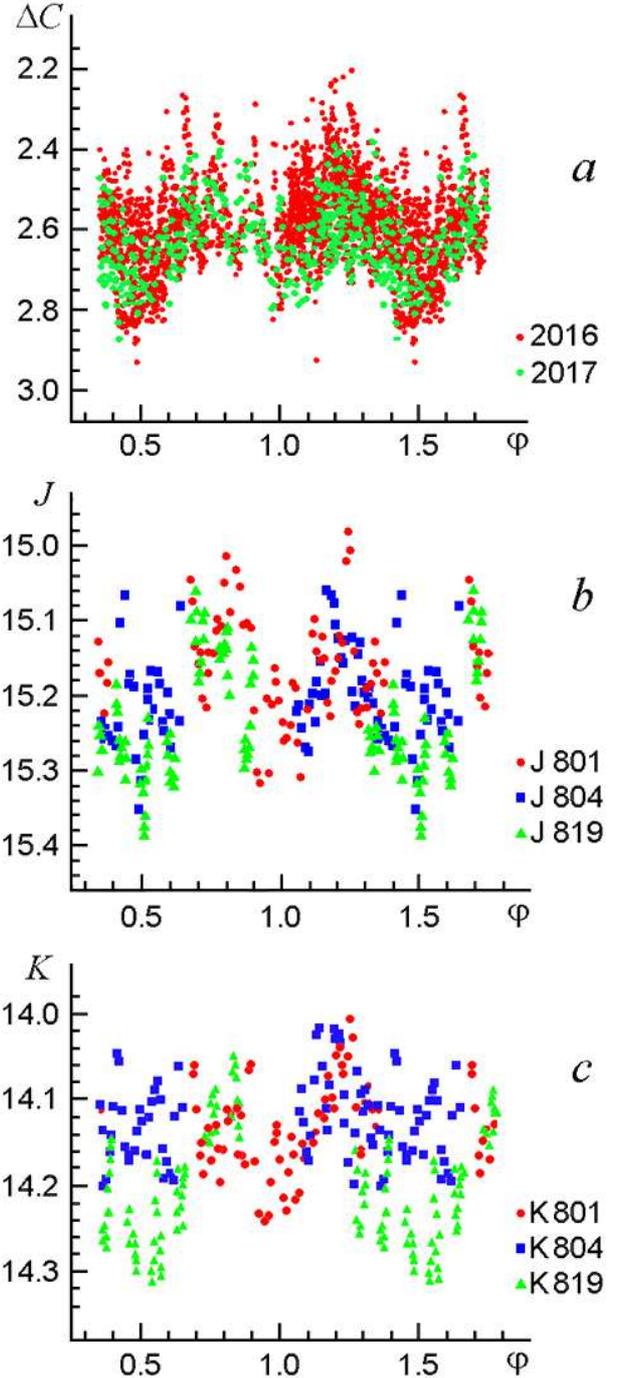}
 \caption{Photometric data on A0620-00 convolved with orbital period. From top to bottom: observation in ''Clear filter'' in 2016--2017, in $J$ and $K$ bands in 2017. The system was in active stage.}
 \label{fig3}
\end{figure}

\E In Figure~\ref{fig3} the measurements for the season 2016--17 are presented convolved with the same orbital period. The double wave being a characteristic feature of the optical star ellipticity effect is also well seen, but, unlikely the first season, the primary and secondary maxima are roughly equal in height, while the $\varphi=1$ minimum is shallower than that of $\varphi=0.5$. Comparing the curve with that of the 2015--16 season, we see that the irregular variability strength (flickering) has increased drastically. The mean system brightness has also increased in all bands: by 0.2~mag in the optical, by 0.25~mag in $J$ and by 0.3~mag in $K$-band which is probably related to an increased contribution of the non-stationary accretion disc and synchrotron radiation of relativistic jets in the system total flux in 2017. The infrared color indices for the orbit-averaged brightness are  $J-K=0.79$ in 2016 and 0.84 in 2017, i.e. the flickering and mean flux growth is accompanied by some little system reddening probably related to the synchrotron radiation impact. So, to judge by all the signs (see e.g. \citealt{cantrell2010,casares_jonker2014}) in 2016--2017 we evidenced the A0620-00 system transiting from passive to active stage of quiescence that happened in less than 230 days. Remarkably, the relative flickering amplitude in A0620-00 (in magnitudes) decreases only slightly from the optical to the infrared.

\section{Mean light curve simulations}\label{sec:modelling}


\E Fast irregular variability (flickering) has a character of flares which are imposed upon the regular orbital variability \citep{Shahbaz}. Authors recommend to extract the regular orbital light curve as a lower envelope of observed points which reflects better the optical star and accretion disc contribution into the total brightness of the system. Since a lower envelope outlining is hard to formalize, we decided to restrict ourselves to calculation of mean observed light curves for A0620-00. These curves are obtained by a {\em group averaging} when single brightness estimates similar in phase and corresponding phase values are averaged to provide a smooth resulting curve which serves to restore model parameters (see below).


\E Such an approach being adopted, the mean squared rms amplitude of deviations of observed values from a ``theoretical'' (i.e. model) light curve defines the mean flickering amplitude plus some observational error biasing. For objective comparison of the flickering amplitude in optical and infrared ranges this seems well appropriate.


\E Assuming normal distribution law for deviations of individual observed points from their arithmetic mean, we fit the model light curve to the so derived mean observed light curve. The residual deviations of individual points from the model light curve, corrected for contribution of measurement errors (that are much lower than the flickering amplitude), define the irregular physical variability of the system due to flickering. Due to flare-type character of flickering, we thus get an underestimated value of the flickering amplitude. As we are mainly interested not in the absolute flickering amplitude but in its dependence on activity stage and wavelength so such a simplified (but free of subjectivity in contrast to lower envelope outlining) analysis procedure appears satisfactory.

\E Mean A0620-00 light curves interpretation was performed in terms of two models: the model with spots on the optical star and without spots.
\begin{enumerate}
\E \item The model without spots on the optical star describes light curve features by existence of a ``hotline'' near the outer (lateral) surface of the accretion disc, at the front (windward) side of the gas stream falling through the Lagrange point $L1$ \citep{khruzina2011}. A physical foundation of such a model is created by 3D gas-dynamical calculations of gas streams in close binary systems \citep{bisikalo}.
\E \item The model with a spotty optical star \citep{Khruzina_cher1995} apart from the ``hotline'' provides an existence of ``cold'' spots on the optical companion surface.
\end{enumerate}


\E It is assumed that the system consists of an optical star filling its Roche lobe which relative size is determined by the components mass ratio $q = {M_\textrm{X}}/{M_\textrm{V}}$, where $M_\textrm{X}$ and $M_\textrm{V}$ are the relativistic and optical component masses, respectively. Fluxes from the optical star surface are computed with an account of gravitational darkening and non-linear limb darkening law.


\E The relativistic object is modelled by a sphere with a small radius $R_1$. It is surrounded by an elliptical accretion disc ($e \la 0.2-0.3$) with a semi-major axis $a = R_\textrm{d}$ and orientation angle $\alpha_\textrm{e}$ (between the disc periastron direction and line connecting component centres). The disc is optically thick and has a complex shape being geometrically thin in the neighborhood of the central object and thick at the outer edge with an opening angle  $\beta_\textrm{d}$. The temperature radial distribution is described by $T(r) = T_{\textrm{in}}(R_1/r)^{\alpha_\textrm{g}}$, where the parameter $\alpha_{\textrm{g}} < 0.75$. The elementary area temperature is computed with an account of the optical star illumination (being insignificant in our case) and warming from high-temperature radiation coming from internals of the disc.


\E A {\em hotline} (HL) adjoins the disc; there is also a {\em hotspot} (HS) at the disc lateral side, at the leeward side of HL.  HL is a shock wave in the narrow region along the gas stream edge which is induced by the infalling stream gas interaction with the corotating near-disc halo. In the model, the radiating part of HL is approximated by a truncated ellipsoid while the HS -- by a semi-ellipse located at the lateral disc side, backward of the stream. The ellipse centre coincides with the point where the gas stream axis crosses the disc geometric body.


\E The novel idea that a gas stream may penetrate into the accretion disc and interact with its material by lateral ({\em\!wall}) shock instead of front shock was advanced by Chochol et al. on the basis of observations analysis \citep{chochol}.


\E All the mentioned structural components radiate as black bodies; the mutual and self-eclipses are taken into account. The detailed description of the parametric model was given in a paper of \citet{khruzina2011}.


\E In the case of the model with spots, this component is added to the model structure as two dark round entities on the surface of a tidally deformed star with a constant temperature contrast across each spot. The contrast is determined by the parameter $F_{\textrm{sp}} = T_{\textrm{sp}}/T_{\textrm{star}}$, where $T_{\textrm{sp}}$ is a spot material temperature and $T_{\textrm{star}}$ is a local temperature of the surrounding stellar atmosphere ($F_{\textrm{sp}} < 1$). A condition for an elementary area on the stellar surface to belong to a spot is the following inequality:
\begin{equation*}
 (x-x_{\textrm{sp}})^2 + (y-y_{\textrm{sp}})^2 + (z-z_{\textrm{sp}})^2 < R^2_{\textrm{sp}},
\end{equation*}
\E where $x, y, z$ are area centre coordinates on the star surface;  $x_{\textrm{sp}}, y_{\textrm{sp}}, z_{\textrm{sp}}$ -- spot centre coordinates, $R_{\textrm{sp}}$ --
its radius. Note that radii and temperatures of spots may be different.


\E Since there are no eclipses observed for this object, the relativistic jets are not considered as distinct components of the model. It is assumed for simplicity that jets just add to the accretion disc brightness.


\E To solve the inverse problem the downhill method \citep{himmelblau} is exploited. As the search for a global minimum of residuals for each light curve proceeded, several tens of initial approximations were tested, because of the known tendency of the residual functional to have many local minima when the number of parameters is high (above 20). As a merit function, the sum of squared deviations of the observed light curve points from the model curve is used; the global $\chi^2$ criterion was used to select the best solution.

\section{Simulation results}\label{sec:mod_results}

\E Since the basic A0620-00 system parameters were firmly established in the paper of \citet{cantrell2010} by a sample of thoroughly selected pattern light curves in passive stage, we did not seek to determine all the system parameters from our light curves. The principal task of our light curves modelling was to extract parameters of a putative spot structure on the optical star surface and to establish a correlation between the accretion disc luminosity (with HL and jets impacts) and the flickering parameters.

\E The total number of model parameters is quite high: 20 parameters for the spotless model and another 8 are added to describe two spots on the optical star. Such a high number of parameters is an advantage, since --- by minimizing the residuals $O-C$ --- it allows to wipe out the impact of systematic errors and to bring the reduced $\chi^2$ value close to unity. This way our multi-parametric model proves to be adequate to observational data and we may be confident that residuals $O-C$ characterize (at the level of random measurement errors) the parameters solely of flickering. What makes the situation easier is that the basic system parameters and, most important, the orbit inclination  $i = 51\degr \pm 0\fdg9$, are firmly established by \citet{cantrell2010}. This constrains some parameters to vary only within reasonable limiting values in our difficult case of an eclipse-free configuration.


\E In particular, orbit inclination was allowed to vary within $i=49\degr-53\degr$, which surrounds the value $i=51\degr$ \citep{cantrell2010}. The mean temperature of the optical star surface may be estimated from its spectral type K3--7V.  \citet{Froning2007} classify the A0620-00 as K5V while estimates made by \citet{oke}, \citet{harrison} range within K3--K7V. The photometric data analysis carried out by \citet{gelino} with no account of the disc contribution resulted in the value K4V while \citet{GonHer2004} basing on the iron lines analysis derived even earlier type K2--3V. The K3--7V range corresponds to the optical star effective temperatures $T_2$ from 4120 to 4760~K \citep{habets}, the range we allowed the mean model temperature to vary within.


\E Of a special interest is the components mass ratio $q$. It is estimated from rotational broadening of absorption lines in the optical star spectrum \citep{wade1988}. Most frequently, the classical line broadening model \citep{collins} is used where a tidally deformed star in a LMXRB is approximated by a flat circle with a linear limb darkening law and a local non-broadened absorption line profile being constant across the disc. In the works of  \citet{petrov2015}, \citet{antokhina} it is demonstrated that the use of a simple classical line broadening model leads to considerable underestimation of $q$. Authors interpret the rotational line broadening in the framework of a more realistic optical star model which takes into account its ``pear-like'' shape and gravitational and limb darkenings. The relations are given to convert $q$ value derived from a simple model which considers classical rotational line broadening into that yielded by more advanced models.

\begin{figure}
 \includegraphics[width=\columnwidth]{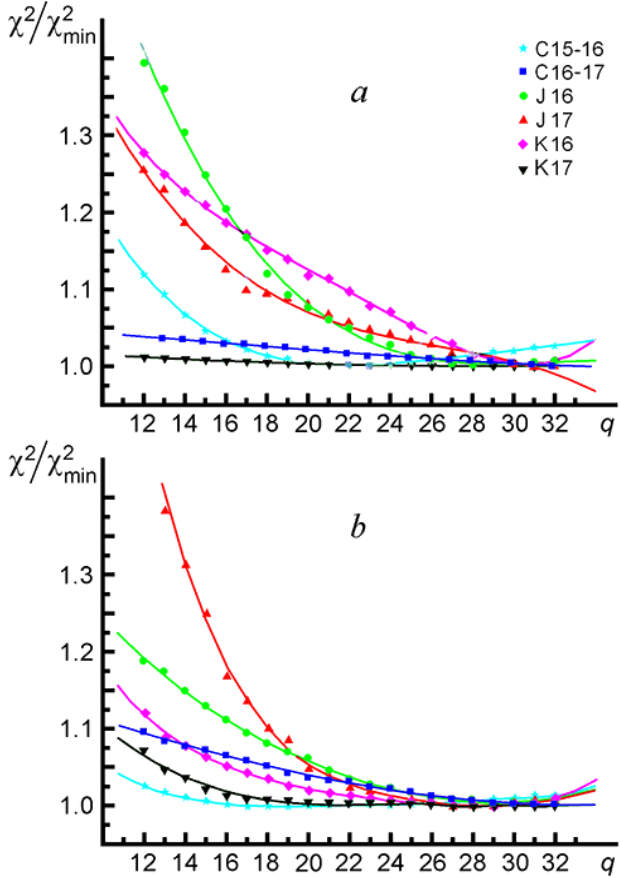}
 \E \caption{Dependence of normalized $\chi^2_\textrm{norm}=\chi^2/\chi^2_\textrm{min}$ values on the mass ratio $q$ derived for the spotless model (a) and for the model with spots (b) for light curves in the bands $J$, $K$ and $C$. Solid lines depict the 3th--4th degree polynomials which approximate the respective calculation data.}
 \label{fig4}
\end{figure}

\E For a number of X ray novae with BHs \citet{petrov2017} quote both the improved  $q$ values obtained on the basis of this more realistic model and optical star masses. For A0620-00 the authors give an improved value $q=26$ instead of the previous $q\sim17$. We thus let $q$ to vary within the range $q=12-32$.  Since we have constrained orbital inclination to be within $i=49\degr-53\degr$ range, some minimal residuals dependence on $q$ arises both for the model with spots and without those (see Figure~\ref{fig4}a,b).

\E From this relation it follows that higher values $q\gid25$ are of preference compared to the approximate value $q\sim17$ derived from the classical line broadening model. It is worth to note that the LMXRB components structure (cold spot parameters, accretion disc luminosity, ``hotline'' and hotspot characteristics) depends on the concrete $q$ value weakly if $q$ is high enough ($>15$).


\E The compact object radius $R_1$ is formally selected to be small enough: $R_1=(0.0001-0.0007)\xi$, where $\xi$ is a distance of Lagrange point $L_1$ from the relativistic object; such a small radius allows to consider it point-like. Accretion disc semi-major axis $a=R_\textrm{d}$, its eccentricity $e$, disc orientation angle $\alpha_\textrm{e}$ and the parameter $\alpha_\textrm{g}$ which sets the law of temperature distribution along the disc radius were adjusted within the following ranges: $R_\textrm{d}/\xi\sim0.2-0.8$, $e\sim0-0.3$, $\alpha_\textrm{e}\sim0\degr-180\degr$ and $\alpha_\textrm{g}\sim0.01-0.75$. Besides, the HL and HS parameters were fitted.


\E For each wavelength range the fit was made by minimizing the $O-C$ discrepancy in the sense of global minimum. All parameters except for $q$ were set free. The minimal residuals depending on a set of $q$ values are plotted in Figure~\ref{fig4}(a,b) in units of minimal residual for each photometric band.


\E From Figure~\ref{fig4} it follows that for large mass ratios ($q>25$) all the light curves (both optical and IR) are equally well fitted both by the model suggesting a spotty optical star and by the spot-free one. Besides, for these $q$ values minimal relative residuals for all wavelengths and observational epochs differ quite a little. So both models provide us with theoretical light curves to compute residual deviations which characterize the flickering amplitude.

\begin{figure}
 \includegraphics[width=\columnwidth]{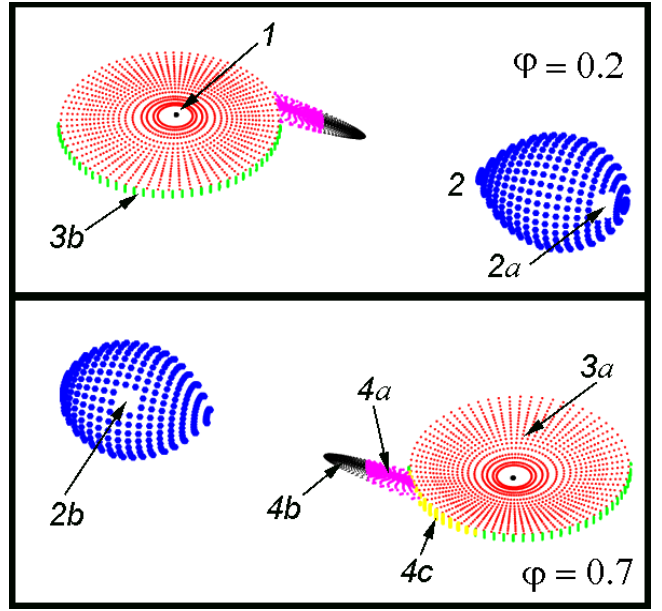}
 \E \caption{Schematic A0620-00 images rendered from the system model with optimal $\Delta C16-17$ light curve parameters for orbital phases $\varphi=0.2$ and 0.7.  Designations: relativistic object (1), red dwarf (2) with two spots on its surface (2a, 2b).  Accretion disc inner (3a) and lateral (3b) sides are shown with the hotspot (4c) backward of the gas stream. (4a) and (4b) are warmed and cold parts of the ``hotline'', respectively.}
 \label{fig5}
\end{figure}

\E Figure~\ref{fig5} gives a model presentation of the A0620-00 binary system (the case of a spotty optical star) in two orbital phases. The scheme corresponds to the parameters found for the optimal optical light curve $\Delta C16-17$ resulting from the minimal $O-C$.

\begin{figure}
 \includegraphics[width=\columnwidth]{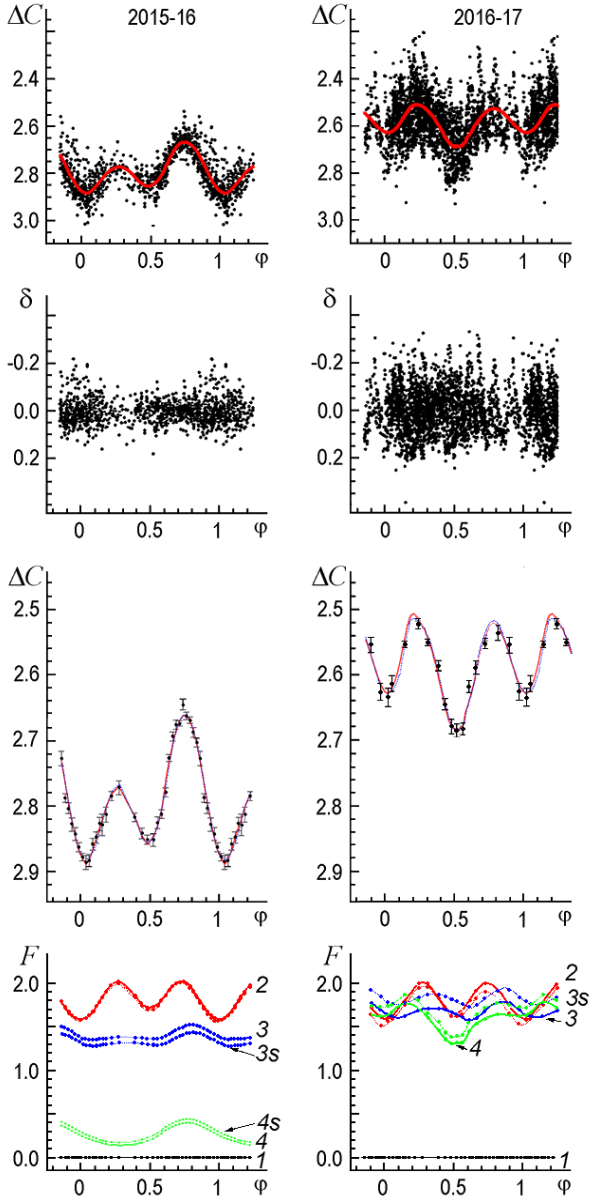}
 \E \caption{Inverse problem solution results interpreting optical light curves $\Delta C15-16$ (left) and $\Delta C16-17$ (right). From top to bottom: individual magnitude differences convolved with the orbital period and the model light curves overplotted for a spot-less case; respective $O-C$ deviations $\delta$; group averaged light curves with overplotted model curves (the model with spots -- red solid curve; without spots -- blue dashed curve);  and the system components flux contribution $F$ in arbitrary intensity units: 1 -- compact object, 2 -- red dwarf, 3 -- accretion disc with hotspot, 4 -- ``hotline''. Suffix $s$ denotes the components of the model with spots.}
 \label{fig6}
\end{figure}

\E Figure~\ref{fig6} shows the inverse problem solutions for the optical light curves of A0620-00 obtained in 2015--16 (passive stage) and in 2016--17 (active stage). In the plot one can see the individual magnitude differences $\Delta C$, folded with the orbital period, and the respective spot-free model light curves which are almost identical to those yielded by the model with spots. The $O-C$ differences $\delta$ are shown in the second graph. Two bottom graphs present mean light curves obtained by group averaging (see Section~\ref{sec:modelling}) with respective rms spread bars and simulated light curves and, below, the relative intensity contribution (in arbitrary units) of different components into the system total brightness. These contributions are shown for models with spots and without spots separately.

\E Since the photometric measurement errors are much less than the flickering amplitudes, Figure~\ref{fig6} demonstrates the A0620-00 system transition from passive into active stage as it is observed in the optical: the mean system brightness rises by about 0.2~mag while the flickering amplitude grows more than twice and becomes comparable with the regular orbital variability amplitude. As it was mentioned above, the orbital curve shape has also changed radically showing equalized maxima and shallower phase $\varphi=0$ minimum in active stage.

\E It is informative to trace the change of each component contribution in different models while the system switches between passive and active stages (Figure~\ref{fig6}, bottom). Due to absence of eclipses, the components fine structure is reconstructed from orbital light curves not that reliably. Simulations yield accretion disc with an eccentricity $e=0.1-0.3$ and a semi-major axis $a=(0.2-0.4)\xi$. Most convincing evidences are obtained for relative component luminosities (optical star, disc, HL):
\begin{enumerate}
  \E \item \textbf{Model without spots on the optical star.} The dwarf optical light contribution is comparable with that of the accretion disc (plus jets), both in passive and in active stages. Meanwhile, the HL flux rises quite significantly and in active stage nearly equals the disc flux, so HL is responsible for the overall system luminosity growth. Since HL is the principal region for the gas stream material interaction with the rotating disc halo (where the shock wave persists), non-stationary processes in this region may lead to the observed intensified flickering in active stage.
  \item \textbf{Model with spots.} In the optical range, the contribution of the disc and HL in passive stage is smaller than that of the optical star. In active stage the system luminosity grows due, for the most part, to significant brightening of the accretion disc which dominates the optical star and also due to substantial increase in the HL brightness. All this taken together (with a possible contribution from relativistic jets) may lead to the intensified flickering in active stage.
\end{enumerate}


\E As stated above, the relatively small contribution of the disc and HL luminosity in passive stage is an explanation for the low flickering amplitude. The ellipsoidal variability anomalies (inequality of maxima) may be simulated both by HL and HS on the disc and by the presence of dark spots on the star surface. The required contrast of spots is 0.6--0.9 for different spectral ranges and the spot sizes are quite considerable: from 0.2 to 0.4 of the star radius $R_2$ (see Figure~\ref{fig5}). For example (see Figure~\ref{fig6}, \ref{fig8}, \ref{fig9}), in 2016 (passive stage) the spots reduced the total $C$ and $J$ luminosity of the system in the first quadrature ($\varphi=0.25$), while in 2017 (active stage) they lowered the $K$ brightness in the third quadrature ($\varphi=0.75$). The spot sizes and location restored by independent modelling of the light curves obtained in different photometric bands coincide within reasonable errors.

\begin{figure}
 \includegraphics[width=\columnwidth]{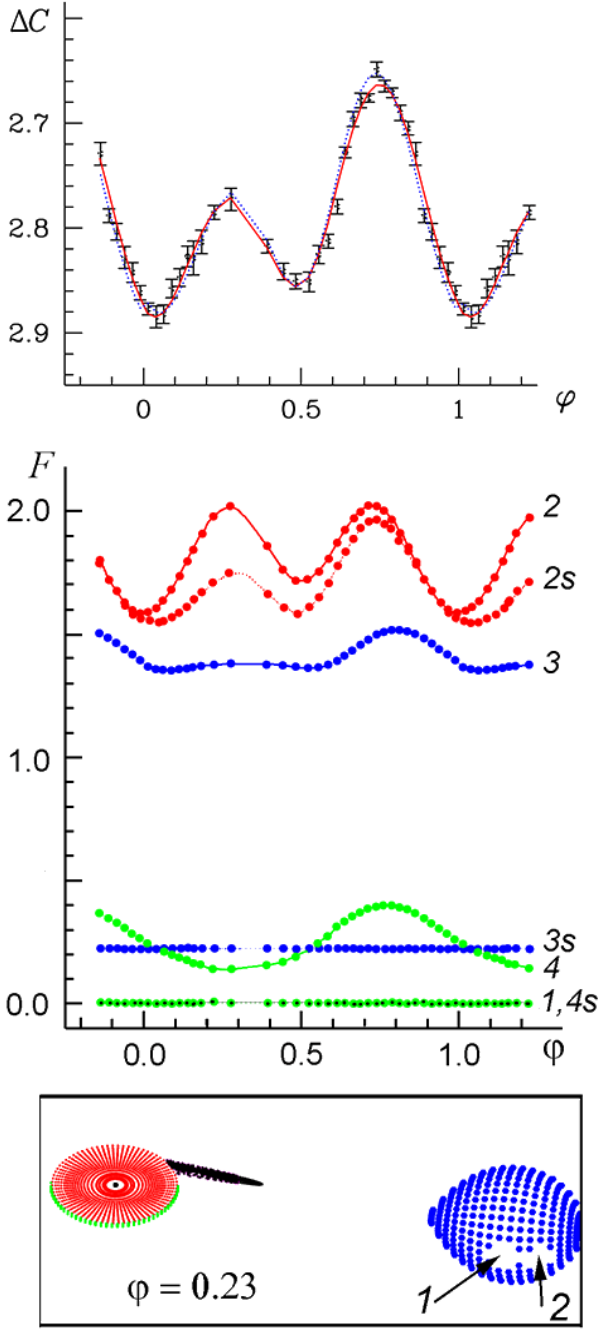}
 \E \caption{Interpretation of A0620-00 optical light curve in passive stage for two model cases. Upper panel: red solid line denotes the optimal curve synthesized for the model with a spotless optical star but with HL and HS; blue dashed line corresponds to the model with a spotty star and with HL and HS ``switched off''. Middle panel: designations for the components contribution are the same as in Fig.~\ref{fig6}. Bottom panel: the ``spotty'' model is rendered for the orbital phase $\varphi=0.23$.}
 \label{fig7}
\end{figure}
\begin{figure}
 \includegraphics[width=\columnwidth]{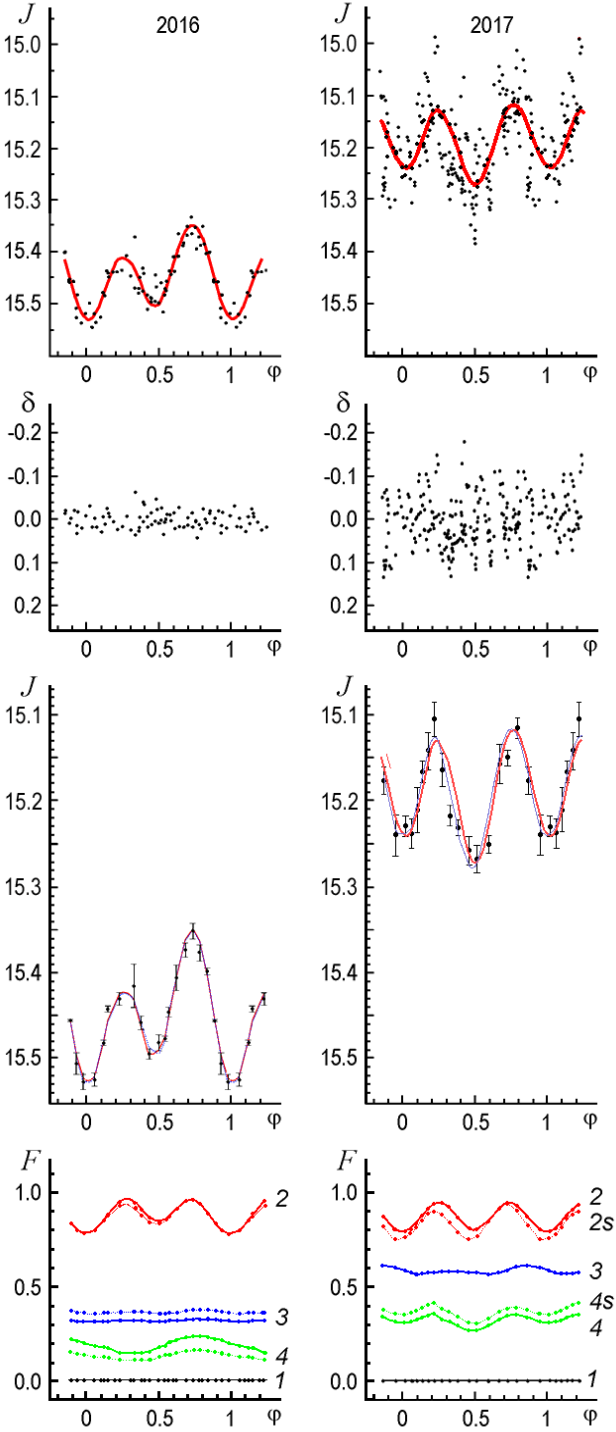}
 \caption{The same as Figure~\ref{fig6} but for $J16$ and $J17$ light curves.}
 \label{fig8}
\end{figure}
\begin{figure}
 \includegraphics[width=\columnwidth]{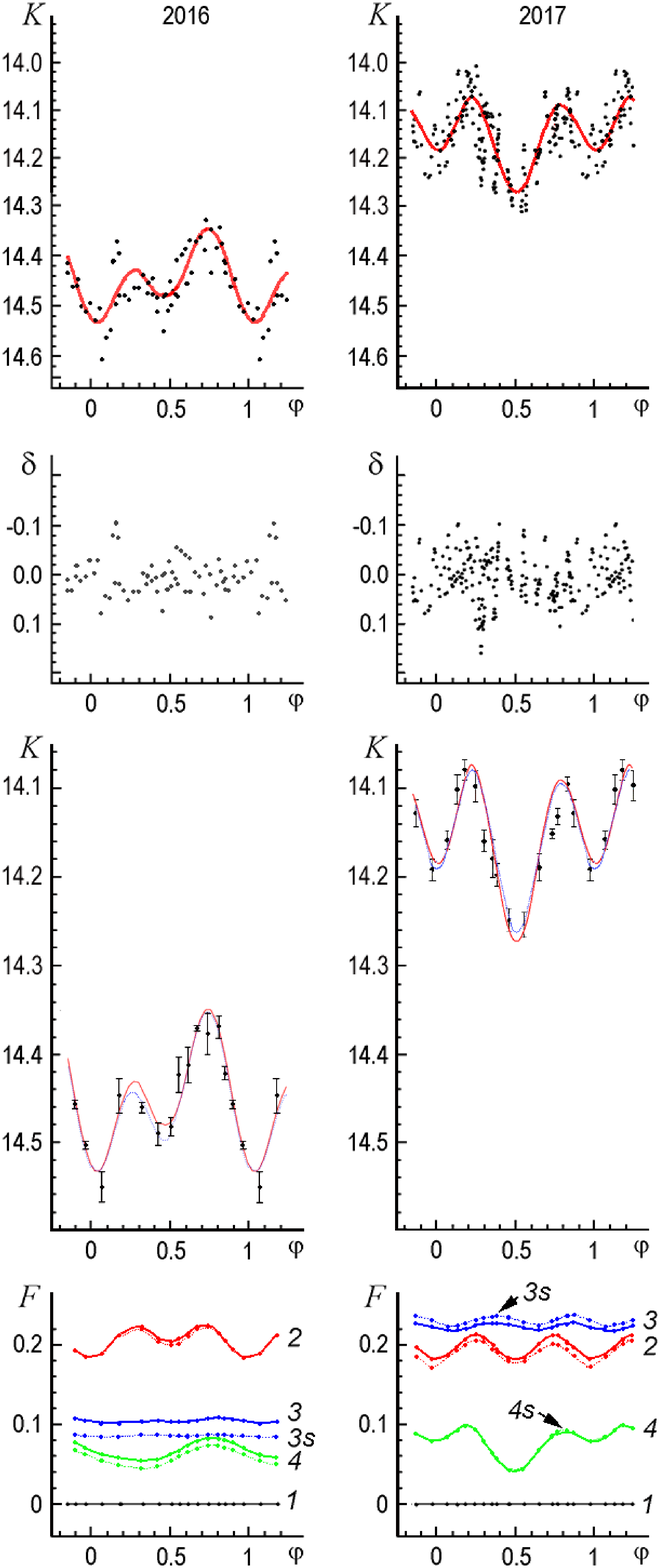}
 \caption{The same as Figure~\ref{fig6} but for $K16$ and $K17$ light curves.}
 \label{fig9}
\end{figure}

\E In order to study the effect of star spots on the orbital light curve in more detail, we have elaborated on two special cases of the model for passive light curve $\Delta C15-16$: with spots but without HL and HS at all, and vice versa -- with HL and HS but without spots. The results are displayed in Figure~\ref{fig7} where the solid line (upper panel) represents the former case and the dashed line is for the latter.

\E It is evident that both special models describe the observed light curve satisfactorily. The spotty model is fully consistent in explanation of the maxima inequality. As expected, the disc contribution becomes independent on phase and observed inequality of maxima is fully reproduced by spot structure. The required relative spot radii $R_\textrm{sp}\cong0.25$ are nearly equal for two closely located spots; their contrast is $F_\textrm{sp}\cong0.9$. A spotless star gives an equal maxima light curve while HL and HS are in charge of the observed maxima inequality.


\E In Figure~\ref{fig8}, \ref{fig9} the inverse problem solution is given for IR light curves. The main properties of optical light curves mentioned above are well reproduced in the near IR: the $J,K$ curves for passive stage (season 2016) demonstrate a double orbital wave with unequal maxima and a deeper primary minimum. The system having switched into active stage, the flickering amplitude grows and both maxima become nearly equal, while $\varphi= 0$ minimum becomes shallower.

\E Both models (with and without spots) predict for the $J$ band that in both stages the optical star dominates, while the light fraction from the accretion disc and HL rises from 2016 (passive stage) to 2017 (active stage).

\E As to the $K$ band, both models coincide in that the optical star dominates in passive stage and at the time the disc contribution is almost equal to that of HL. For active stage both models consider the accretion disc contribution to increase significantly and to tend to be similar with that of the optical star.


\E Consequently, the given models are fit to reproduce, in a satisfactory qualitative manner, basic features of optical and near-IR light curves for A0620-00 both in passive and active stages of quiescence. The ''spotty model'' reproduces the photometric measurements not worse than the classic one without spots (see Figure~\ref{fig7}). That's why the presence of spots on the optical star surface can not be ruled out. As the spots size is rather large ($\sim(0.2-0.4)R_2$) and the magnetic field in spots on the surface of red dwarves is usually of the order of many kilogauss (see e.g. \citealt{gershberg}), this consideration may invest to a plausible explanation of anomalously fast orbital period shortening (\citeauthor{GonHer2017}\citeyear{GonHer2014,GonHer2017}) being in an accord with the standard evolutionary scenario for LMXRBs \citep{Iben}, which attributes the system angular momentum loss with magnetised stellar wind and gravitational wave radiation. Although, it should be noticed that the same orbital light curves may be well reproduced by spotless models as well.

\E \section{Residual deviations analysis and the A0620-00 flickering characterization}
\label{sec:6}


\E Let us explore the dependence of flickering rms amplitude $\Delta F_\textrm{fl}(\lambda)$ (absolute value, i.e. in fluxes) on wavelength $\lambda$. Our photometric observations of A0620-00 cover a considerable wavelength range from $\lambda_\textrm{eff}\sim6400$~{\AA} to 2.2~\micron, so such a  $\Delta F_\textrm{fl}(\lambda)$ relation bears important information on the flickering nature.

\begin{figure*}
 \includegraphics[width=\textwidth]{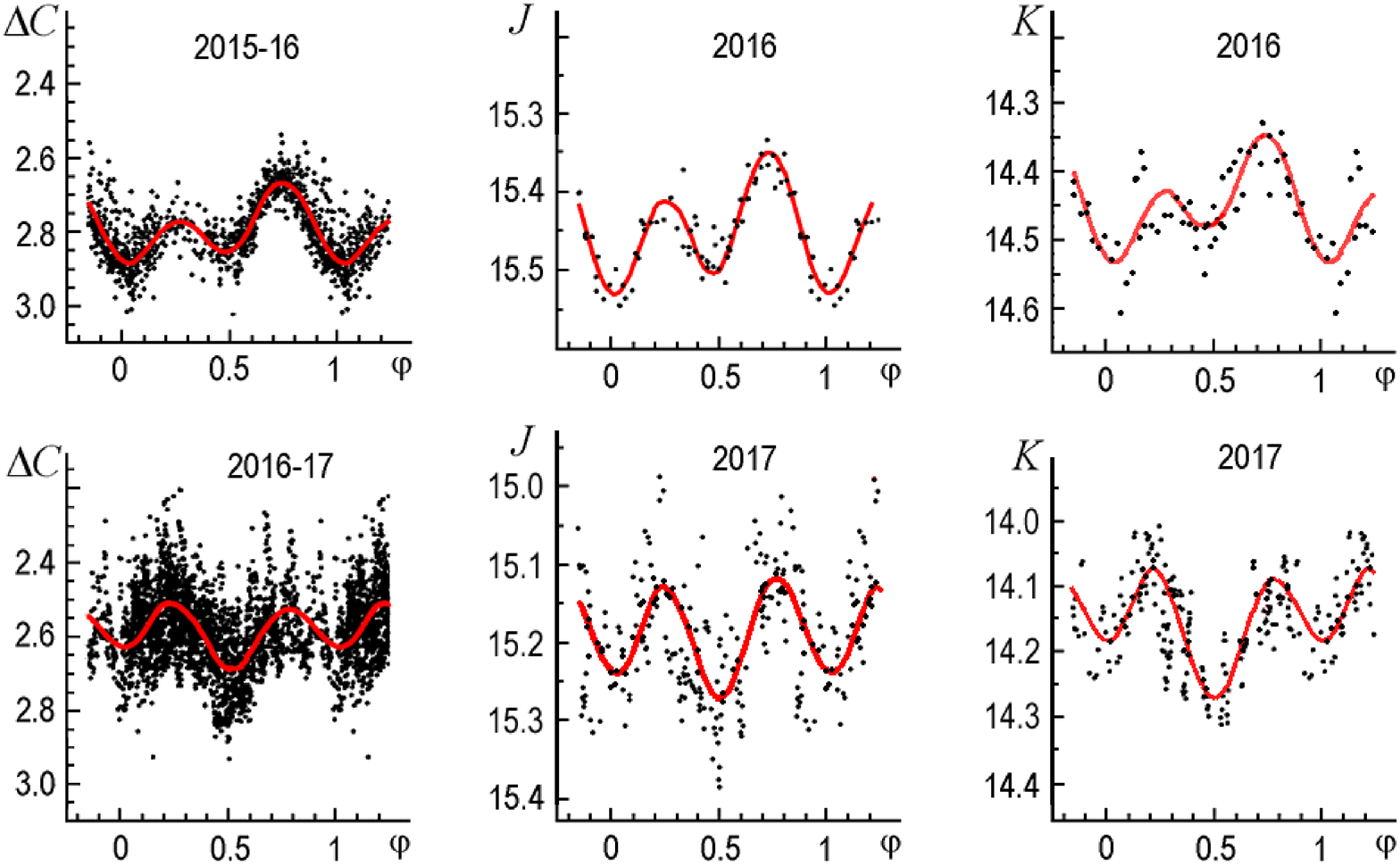}
 \caption{Theoretical (red line) and observational (dots) light curves for A0620-00 in passive (upper row) and in active (lower row) stages.}
 \label{fig10}
\end{figure*}

\begin{figure*}
 \includegraphics[width=\textwidth]{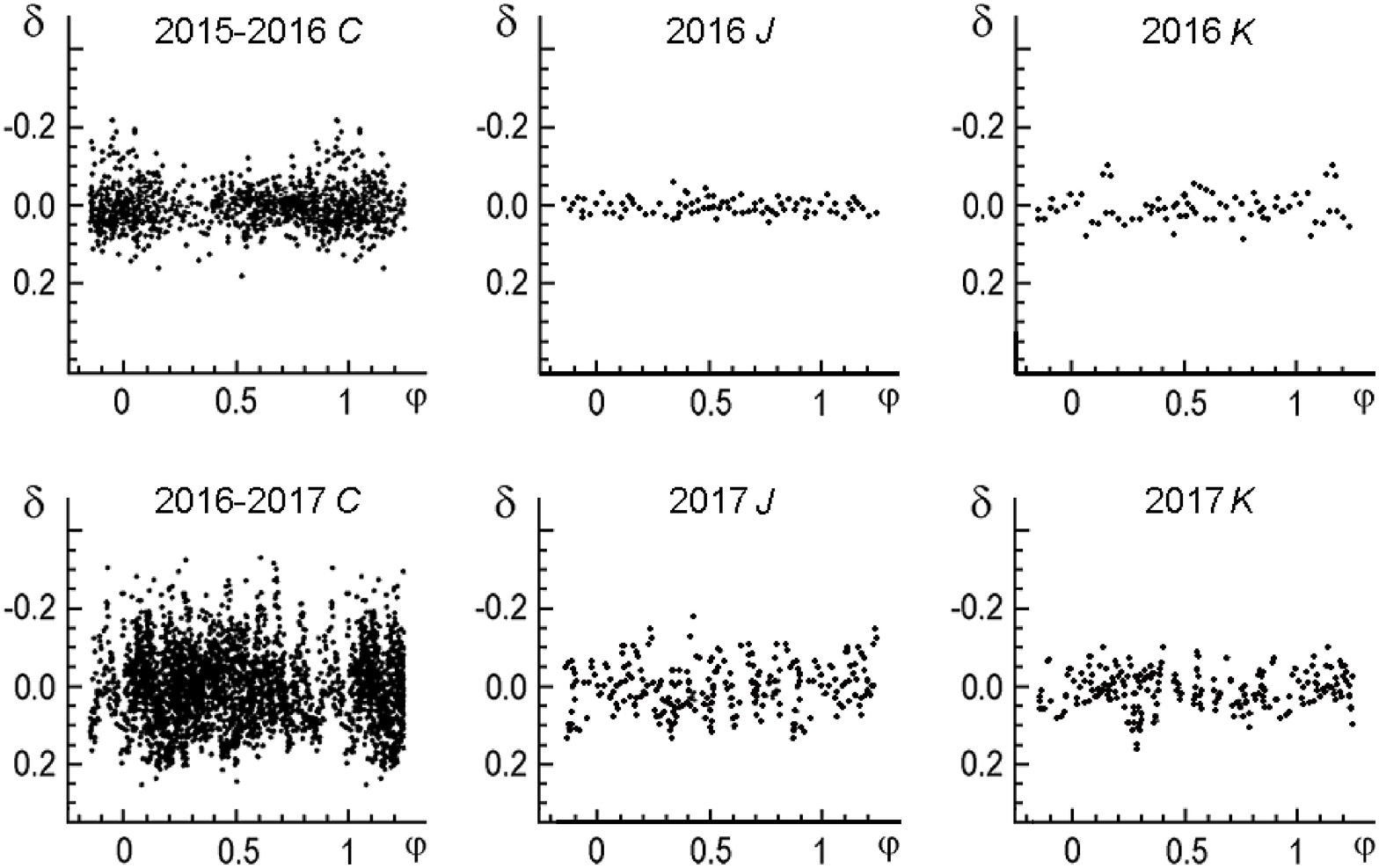}
 \caption{Residual deviations $\delta=m^\textrm{obs}-m^\textrm{t}$ of observed magnitudes of A0620-00 from theoretical light curves, in passive (upper row) and in active (lower row) stage.}
 \label{fig11}
\end{figure*}


\E In Figure~\ref{fig10} we collect observed and model light curves of A0620-00 for both passive and active stages, having folded them with the orbital period, while in Figure~\ref{fig11} the residual deviations $\delta$ are given, all in stellar magnitudes for both optical and infrared bands. Those deviations include observation error (around 0.01--0.02~mag, see Section~\ref{sec:observations}) and real physical variability (flickering). In all cases the deviations are well above the measurement error, so these plots clearly illustrate the flickering characteristics as a function of orbital phase, wavelength and the activity stage of A0620-00.


\E From Figure~\ref{fig11} one can see that flickering amplitude is nearly constant across the orbital period. A possible exclusion may be noticed in passive stage where optical flickering has a tendency to increase near the phase $\varphi=0$ (X-ray source is behind the optical star).


\E A strong flickering amplitude dependence on the activity stage is evident. Expressed in logarithmic units (and thus acting as {\em relative} flickering intensity), flickering magnitude in active stage is twice that in the passive in all photometric bands. In active stage flickering, being most prominent in optical range, is comparable with the regular orbital variability amplitude. In $J, K$ it is weaker but still well noticeable. In passive stage flickering magnitude is the largest in the optical, the smallest in $J$.

\E Since the absolute system flux is different in different photometric bands, the flickering amplitude should be converted into the same units before comparison. In order to do this, we computed absolute spectral flux densities $F_\lambda$ for the A0620-00 system taking into account the observed optical and IR magnitudes corrected for interstellar (IS) extinction and using the absolute $R, J, K$ calibration:
\begin{equation}
 F_\lambda = F_\lambda^0\times2.512^{(m_0-m)},
\end{equation}
\E where $m_0 = 0$, $m$ is the observed magnitude, corrected for IS extinction, $F_\lambda^0$ -- the absolute spectral flux density from a zero magnitude star outside the earth atmosphere. For the optical $R$ band we used the flux calibration from \citet{bessell}: $F_{R}^0 = 1.74\times10^{-10}$~erg~cm$^{-2}$~s$^{-1}$~\AA$^{-1}$, infrared fluxes were taken from \citet{koornneef}: $F_\textrm{J}^0 = 3.14\times10^{-10}$, $F_\textrm{K}^0 = 4.12\times10^{-11}$ in units erg~cm$^{-2}s^{-1}$\AA$^{-1}$. The $F_\lambda$ values derived in this way were corrected for IS extinction with  $A_\textrm{V}\cong1.2$~mag, see \citet{gelino}; the IS extinction law from \citet{rieke} was used.

\E Since the orbital variability amplitude is relatively small, we restricted ourselves to calculation of absolute flux densities averaged over the whole orbital period $\bar{F}_\lambda$.

\begin{figure}
 \includegraphics[width=\columnwidth]{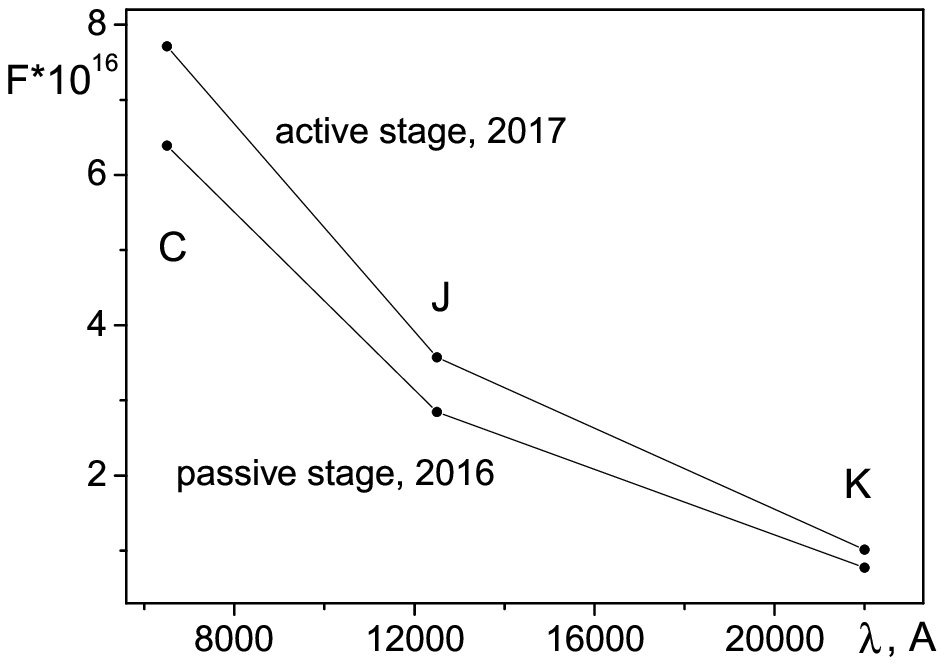}
 \caption{Orbital period mean flux density for A0620-00 in active (upper points) and passive (lower points) stage. Flux density is corrected for interstellar extinction. The scale is $\bar{F}_\lambda\times10^{16}$~erg~cm$^{-2}$~s$^{-1}$~\AA$^{-1}$.}
 \label{fig12}
\end{figure}


\E In Figure~\ref{fig12} and in Table~\ref{tab2} the absolute period-averaged flux densities for the A0620-00 system corrected for IS extinction are given as functions of wavelength $\lambda$. Since the central wavelength of the ''white `` $C$ band is $\lambda_\textrm{eff}\sim6400$~\AA, we used the $R$ magnitude ($\lambda\sim7000$~\AA, the Johnson $UBVRI$ system) to calibrate optical observations. Given the wide span of wavelengths under consideration ($\lambda\lambda6400-22000$~\AA), such a substitution of $C$ by $R$ is appropriate for our aims. As it is evident from Figure~\ref{fig12}, the absolute system flux corrected for IS extinction drops by a factor of 8 for as we pass from $\lambda=6400$~\AA\ to $\lambda=22000$~\AA; at the same time, the mean system flux rises by 20--30~per cent from passive to active stage in all bands.

\begin{table*}
\caption{Parameters of flickering for A0620-00}
  \label{tab2}
\begin{threeparttable}
  \begin{tabular}{lcccc}
  \hline
  $\lambda$ & $\bar{m}$ & $\Delta m_\textrm{fl}$ & $\bar{F}$ & $\Delta F_\textrm{fl}$\\
  \AA & mag & mag & erg~cm$^{-2}$~s$^{-1}$~\AA$^{-1}$ & erg~cm$^{-2}$~s$^{-1}$~\AA$^{-1}$ \\
  \hline
  \multicolumn{5}{c}{2016, passive stage} \\

6400 & $16.090\pm0.003$ & $0.045\pm0.001$ & $6.39\times10^{-16}$ & $(2.647\pm0.084)\times10^{-17}$\\
12500 &	$15.109\pm0.003$ & $0.011\pm0.003$ & $2.85\times10^{-16}$ & 	$(0.283\pm0.071)\times10^{-17}$\\
22000 &	$14.316\pm0.005$ & $0.031\pm0.005$ & $0.75\times10^{-16}$ & $(0.222\pm0.034)\times10^{-17}$\\
$J-K$ & 0.793\\	
\multicolumn{5}{c}{2017, active stage} \\

6400 & $15.887\pm0.002$ & $0.093\pm0.002$ & $7.71\times10^{-16}$ & $(6.791\pm0.144)\times10^{-17}$\\
12500 &	$14.862\pm0.004$ & $0.059\pm0.004$ & $3.57\times10^{-16}$ & $(1.985\pm0.139)\times10^{-17}$\\
22000 & $14.024\pm0.004$ & $0.047\pm0.004$ & $1.01\times10^{-16}$ &	$(0.447\pm0.035)\times10^{-17}$\\
$J-K$ & 0.838\\	
\hline
\end{tabular}

\E Notes. $\bar{m}$ -- an orbital period mean magnitude of the system corrected for interstellar extinction, $\Delta m_\textrm{fl}$ -- flickering amplitude (see Equation~\ref{eq3} and \ref{eq5}), $\bar{F}$ -- mean system flux density, $\Delta F_\textrm{fl}$ -- rms flickering amplitude in units of flux density.

\end{threeparttable}
\end{table*}

\begin{figure}
 \includegraphics[width=\columnwidth]{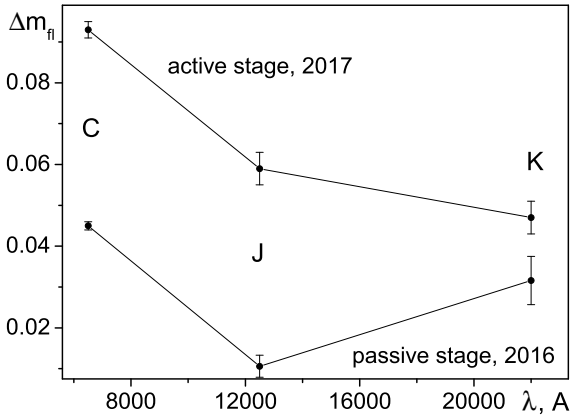}
 \caption{Rms flickering amplitude (in magnitude) versus wavelength for two stages of the A0620-00 system: upper points represent active stage in 2016--2017, lower points correspond to passive stage in 2015--2016. Error bar is equal to the flickering amplitude uncertainty (see Equation~\ref{eq3}, \ref{eq4}, \ref{eq5} and Table~\ref{tab2}).}
 \label{fig13}
\end{figure}

\E In Figure~\ref{fig13} the period-averaged rms flickering amplitude expressed in magnitudes $\Delta m_\textrm{fl}$ is plotted as function of wavelength:
\begin{equation}\label{eq3}
  \Delta m_\textrm{fl}=\sqrt{\Delta m^2-\sigma^2_\textrm{obs}},
\end{equation}
\E where $\sigma_\textrm{obs}=0.01\div0.02$~mag -- the mean measurement error for an individual point on the observed light curve,
\begin{equation}\label{eq4}
  \Delta m=\sqrt{\sum^{N}_{k=1}(m^\textrm{obs}_k-m^\textrm{t}_k)^2/(N-1)}.
\end{equation}
\E Here $m^\textrm{obs}_k$ is the observed system magnitude, $m^\textrm{t}_k$ is the model magnitude, $N$ -- the number of individual measurements in the light curve, $\delta=m^\textrm{obs}_k-m^\textrm{t}_k$. Accuracy of the mean flickering amplitude is
\begin{equation}\label{eq5}
  err=\sqrt{\sum^N_{k=1}(m^\textrm{obs}_k-m^\textrm{t}_k)^2}/N.
\end{equation}
\E Similar formulae allow to compute the respective quantities, namely the rms flickering amplitude and its error, in absolute fluxes $\Delta F_\textrm{fl}(\lambda)$. It is seen that $\Delta m_\textrm{fl}$ in active stage decreases monotonically from $\sim$0.093~mag to 0.047~mag as wavelength grows from 6000 to 22000~\AA. In passive stage $\Delta m_\textrm{fl}$ is significantly lower and lies between 0.045 and 0.011~mag. In this state $\Delta m_\textrm{fl}(\lambda)$ changes with wavelength non-monotonically: it is minimal in the $J$ band  ($\lambda=12500$~\AA) and rises again to the $K$ band ($\lambda=22000$~\AA). Such a feature hints at a possible two-component flickering structure for the A0620-00 system. The thermal component presumably caused by flares in the accretion disc and in the disc-stream interaction region drops sharply from the optical range ($\lambda=6400$~\AA) to the $J$ band while further in the infrared the non-thermal (synchrotron) flickering component starts to be significant, due to its weaker dependence on $\lambda$.

\begin{figure}
 \includegraphics[width=\columnwidth]{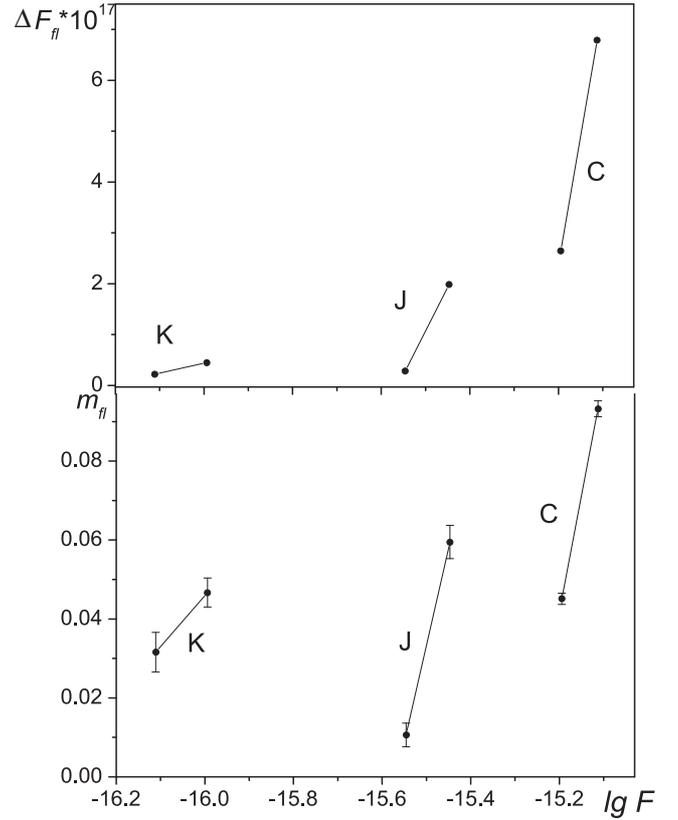}
 \caption{Rms flickering amplitude in different bands versus flux density. Flickering amplitude is expressed in units of flux density (top) and in magnitudes (bottom).}
 \label{fig14}
\end{figure}


\E This idea is illustrated by Figure~\ref{fig14} in which one can see how the rms flickering amplitude in fluxes and in magnitudes grows with the mean system luminosity. While in the optical and $J$ bands the flickering growth curves follow quite the same law and the slope is steep, in the $K$ band the slope is flatter. This may additionally evidence that flickering dependence on the mean system flux in the infrared may be related not only to the disc activity, i.e. HL and HS phenomena, but also to some other reasons, e.g. to non-stationary synchrotron radiation from relativistic jets.

\begin{figure}
 \includegraphics[width=\columnwidth]{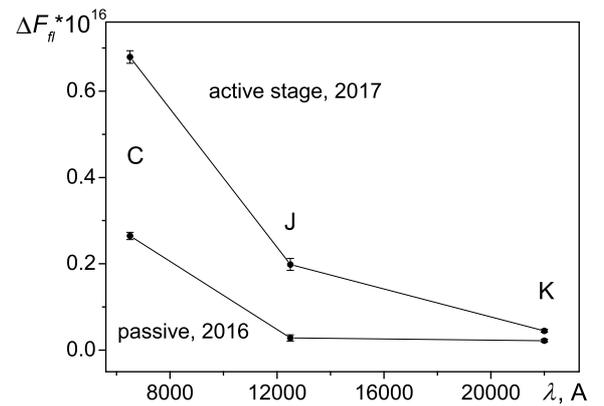}
 \caption{Rms flickering amplitudes expressed in units of flux density and corrected for interstellar extinction versus wavelength. Two stages of the A0620-00 system are plotted: top -- active stage in 2016--2017, bottom -- passive stage in 2015--2016.}
 \label{fig15}
\end{figure}

\E In Figure~\ref{fig15} and in Table~\ref{tab2} the rms flickering amplitudes are given in flux units as functions of wavelength. An {\em approximate} estimate may be obtained from the following formula:
\begin{equation}
  \Delta F_\textrm{fl}(\lambda)\cong \bar{F}_\lambda \Delta m_\textrm{fl}(\lambda),
\end{equation}
\E where $\bar{F}_\lambda$ is a mean absolute flux density of the system, corrected for IS extinction.

\begin{figure}
 \includegraphics[width=\columnwidth]{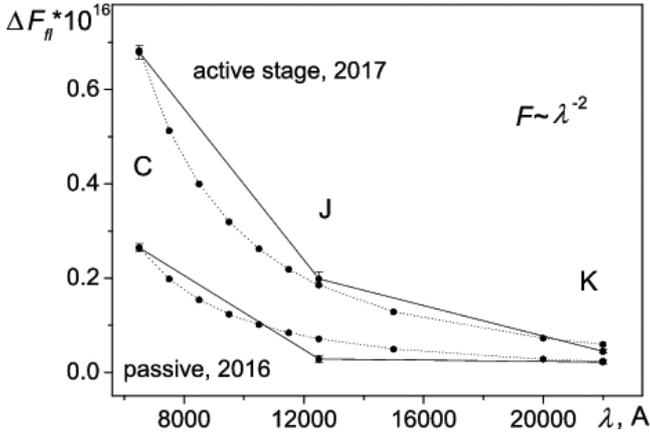}
 \caption{Flickering amplitude fitted by the $\lambda^{-2}$ law for active and passive stages.}
 \label{fig16}
\end{figure}

\E In Figure~\ref{fig16} approximation of observed $\Delta F_\textrm{fl}(\lambda)$ dependences for passive and active stages is given in form of a $\lambda^{-2}$ law. Remarkably, in active stage this law fits observational relation $\Delta F_\textrm{fl}(\lambda)$ quite well. It corresponds to the thermal radiation of optically thin high-temperature plasma ({\em bremsstrahlung}).
\E In passive stage, as follows from Figure~\ref{fig16}, the observed $\Delta F_\textrm{fl}(\lambda)$ relation does not obey the $\lambda^{-2}$ law.

\begin{figure}
 \includegraphics[width=\columnwidth]{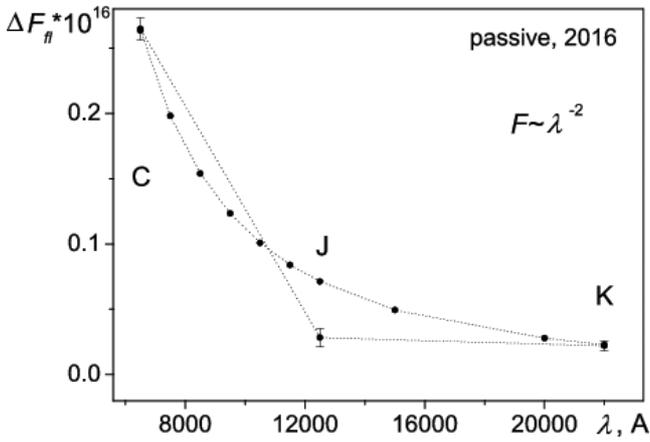}
 \caption{An attempt to fit flickering amplitude by the $\lambda^{-2}$ law for passive stage.}
 \label{fig17}
\end{figure}
\begin{figure}
 \includegraphics[width=\columnwidth]{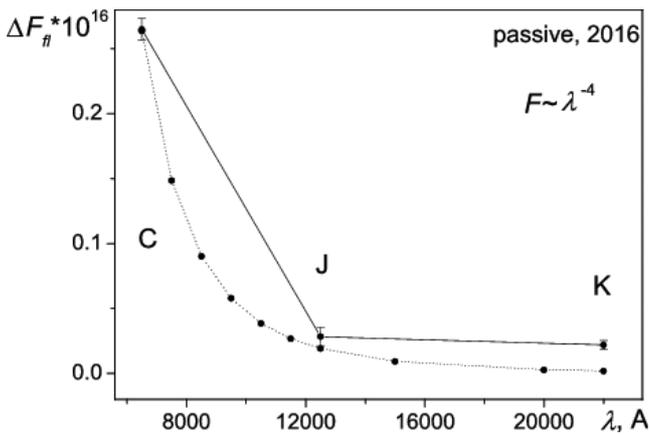}
 \caption{Flickering amplitude fitted by the $\lambda^{-4}$ law for passive stage. A two-component structure of flickering amplitude is seen: flickering amplitude -- wavelength relation is well described by a $\lambda^{-4}$ law in $\lambda\lambda6400-12500$~\AA\ range, and is flat further to $\lambda=22000$~\AA.}
 \label{fig18}
\end{figure}

\E In Figure~\ref{fig17} approximation of the observed dependence $\Delta F_\textrm{fl}(\lambda)$ for passive stage by the $\lambda^{-2}$ law is given separately for clarity. It is evident that all the data throughout the observed range $\lambda\lambda6400\div22000$~\AA\ cannot be fitted by a common law because it is of rather two-component origin (see Fig.~\ref{fig18}).

\E In the short wavelength domain $\lambda\lambda6400-12500$~\AA\ the relation $\Delta F_\textrm{fl}(\lambda)$ is well described by a $\lambda^{-4}$ law, while at longer wavelengths $\lambda\lambda12500-22000$~\AA\ it is flat: $\Delta F_\textrm{fl}(\lambda)\cong \textrm{const}$. The fourth power corresponds to the thermal radiation of optically thick high-temperature plasma (Rayleigh--Jeans spectral domain) while in the flat longer-waves part synchrotron radiation is to appear.


\E It is well known that the synchrotron radiation spectrum obeys a power law and, depending on the particular power exponent in the energy distribution of relativistic electrons, the radiation intensity may decrease with wavelength much slower than the thermal radiation intensity does (see e.g. \citealt{ginzburg}).

\E In case the relativistic electrons in jets are distributed according to the power law
\begin{equation}
  \textrm{d}N/\textrm{d}E\sim E^{-\delta},
\end{equation}
\E then synchrotron radiation in the optically thin case will also have a power-law spectrum:
\begin{equation}
  I_\nu\sim \nu^{-(\delta-1)/2}.
\end{equation}

\E Recalling that $|\textrm{d}\nu|=c/\lambda^2|\textrm{d}\lambda|$, the wavelength distribution of synchrotron radiation is as follows:
\begin{equation}
  I_\lambda\sim \lambda^{(\delta-5)/2}.
\end{equation}

\E If one considers a spectrum of cosmic rays which particles are accelerated at the strong shock wave fronts (first-order Fermy mechanism)
\begin{equation}
  \textrm{d}N/\textrm{d}E\sim E^{-2},
\end{equation}
\E then, taking analogously $\delta=2$, we obtain possible wavelength distribution of synchrotron radiation $I_\lambda$:
\begin{equation}
  I_\lambda\sim\lambda^{-1.5}.
\end{equation}

\E This relation is weaker than that for free--free emission of optically thin high-temperature plasma ($I_\lambda\sim\lambda^{-2}$). Taking a higher exponent in energy distribution for relativistic electrons, e.g. $\delta=4$, we obtain a nearly flat synchrotron radiation spectrum: $I_\lambda\sim\lambda^{-0.5}$.

\E In the paper of \citet{kardashev} various systematic and stochastic mechanisms of relativistic electrons acceleration are considered with respective synchrotron radiation spectra. This also evidences that in the wavelength scale there may be cases of a nearly flat spectrum.

\E The conclusion on the possible existence of a distinct, synchrotron component in the flickering phenomenon of A0620-00 ties well with the recent discovery of the linear polarization ($p = 1.25\pm0.28\%$) of the infrared radiation from this system \citep{russell2016}.


\section{Conclusion}\label{sec:conclusion}

\E We performed the analysis of regular orbital brightness variations of the A0620-00 system in the optical and near-infrared $J,K$ bands and studied the flickering characteristics for this system both in passive and active stages focusing on wavelength dependence in a wide spectral range $\lambda\lambda=6400\div22000$~\AA (i.e. wavelength changes by a factor of 3.4). In time less than 230 days the system, staying in quiescence (X-ray luminosity is low, about $\sim3\times10^{30}$~erg~s$^{-1}$) transited from {\em passive} stage into the {\em active} which was accompanied by about $\sim0.2-0.3$~mag mean system brightness growth in all studied bands while the flickering amplitude increased more than twice. The shape of regular orbital light curves of A0620-00 also changed radically after this transition. The correlation of the mean flux and flickering amplitude conforms to previous findings \citep{cantrell2008,cantrell2010} and evidences the growth of the contribution of the accretion disc, disc-stream interaction region and possibly the relativistic jets radiation in the total luminosity of the system. Thus, despite the fact that the system remains in quiescent state with a very low X-ray luminosity \citep{garcia2001}, there are some violent processes to occur in this state leading to formation of non-stationary relativistic jets.

\E The regular brightness variations of A0620-00 are fitted in the framework of two models for optical star, with and without spots, which reproduce observations equally well. The fact the model with spots does not seem to contradict observations gives some reason to consider the optical component as a magnetically active star, that may be essential in understanding the nature of fast orbital period shortening in this and similar systems \citep{GonHer2014,GonHer2017}.

\E An important result of our modelling of regular orbital light curves for A0620-00 is the ascertainment of correlation between the accretion disc and ``hotline'' (including possible relativistic jets as well) luminosity with the flickering amplitude. In passive stage the contribution of the accretion disc and ``hotline'' is small compared to the optical star luminosity, so the flickering amplitude is small as well. In active stage the disc brightness is comparable with that of the optical star, while the flickering amplitude is doubled.

\E Having calibrated observed fluxes in white light and $J, K$ bands and taking into account interstellar extinction, we calculated the root-mean-square flickering amplitude $\Delta F_\textrm{fl}(\lambda)$ as a function of wavelength for both passive and active stages. This relation bears important information on the flickering nature.

\E Analysis of the $\Delta F_\textrm{fl}(\lambda)$ dependence throughout the $\lambda\lambda6400\div22000$~\AA\ range has pointed out possible existence of two flickering components in A0620-00: thermal and non-thermal, with the latter possibly originating from non-stationary relativistic jets. In active stage the thermal component (bremsstrahlung from the optically thin high-temperature plasma) dominates throughout the studied wavelength domain. In passive stage this component decreases sharply with wavelength and in the longer wave range $\lambda\lambda12500\div22000$~\AA\ the non-stationary (possibly synchrotron) component becomes prominent having a contribution nearly independent on wavelength.

\E This conclusion, on a qualitative level, agrees well with the recent detection of linear infrared light polarization for the A0620-00 system in quiescence \citep{russell2016}. The results which we have obtained are to be confirmed on the basis of more extended observational data, so further photometric monitoring of this system in optical and infrared bands is of great interest.

\section*{Acknowledgements}

\E This work was performed with the equipment purchased from the funds of the Program of Development of Moscow University.
\E The effort of A.M.~Cherepashchuk and N.I.~Shatsky is supported by RNF grant 17-12-01241 (problem formulation, infrared observations and the analysis, physical conclusions). N.A.~Katysheva, T.S.~Khruzina and S.Yu.~Shugarov thank the ``Leading scientific schools'' program NSh-9670.2017.2 for the support (optical observations and computer modeling). S.Yu.~Shugarov thanks for partial support of his research by grants VEGA 2/0008/17 and APPV 15-0458 (optical observations). Authors are grateful to K.A.~Postnov for useful discussions.

\label{lastpage}
\end{document}